\documentstyle[a4p,12pt,cite,epsfig]{article}
\def\RI{$R_{1D}\ $} 
\begin{document} 
\begin{titlepage} 
\vspace{7mm} 
\begin{center} 
{\Large{\bf On the role of the time scale $\Delta t$\\
\vspace{2mm}
in Bose$-$Einstein correlations}} 
\end{center}
\bigskip  
\begin{center} 
 \vspace{8mm} 
{\large\bf Gideon Alexander\footnote{\it gideona@post.tau.ac.il} and  
Erez Reinherz-Aronis\footnote{\it erezra@post.tau.ac.il}} 
\end{center}   
\vspace{2mm}

\centering{\it School of Physics and Astronomy}\\ 
\centering{\it Raymond and Beverly Sackler Faculty of Exact Sciences}\\ 
\centering{\it Tel-Aviv University, 69978 Tel-Aviv, Israel}\\ 
 
\vspace{10mm}  
 
\begin{abstract} 
\vspace{2mm}   
{\small
The time scale parameter $\Delta t$, which appears in the 
Bose-Einstein Correlations (BEC) treated in terms of the Heisenberg
uncertainty relations,
is reexamined. Arguments are given for the role of $\Delta t$ to be
a measure of the particles' emission time rather than representing the
strength property of the correlated particles. Thus in the
analyses of the $Z^0$ hadronic decays, the $\Delta t$ given value of 
$\sim$$10^{-24}$ 
seconds, is the particles' emission time
determined by the $Z^0$ lifetime.
In heavy ion collisions $\Delta t$ measures the emission time
of the particles produced in a nucleus of atomic number $A$.  
This emission time 
is shown here to be equal to $\Delta t =\frac{m_{\pi}a^2}{\hbar c^2}A^{2/3}$
that is, 
proportional to the nucleus surface area. 
This relation agrees rather well with the experimental
$\Delta t$ values deduced from the BEC analyses of heavy ion collisions.}
\end{abstract} 
\vspace{1cm} 
\begin{flushleft} 
\end{flushleft}
\end{titlepage} 
 
\section{Introduction} 

Bose-Einstein Correlations (BEC) of identical pairs of spin zero hadrons, 
produced in a variety of particle interactions, have been studied  
over more than four decades. From those studies which were carried out 
in one dimension, a single length parameter \RI was   
extracted which is taken to represent the radius
of a spherical symmetric Gaussian 
interaction volume. Further on these studies were extended 
to the spin 1/2 baryon pairs, like the proton-proton and the $\Lambda \Lambda$ 
systems, by using the Fermi-Dirac Correlations (FDC) procedure
\cite{review}. 
Experimentally it has been observed 
that in the $Z^0$ gauge boson hadronic decays 
the \RI value decreases as the hadron mass 
$m$ increases \cite{review}. To account for this feature two main approaches have 
been proposed. The first \cite{acl} rests on the Heisenberg uncertainty relations  
which correlate the dimension \RI with $m$ via a time scale parameter 
$\Delta t$. In the same approach one has also considered the
three dimensions (3D) BEC analyses having the axes
$R_{long}$, $R_{side}$ and $R_{out}$, defined in the Longitudinal Center
of Mass System (LCMS) \cite{csorgo}. In this case, similar to the relation of \RI
to $m$ and $\Delta t$,  
$R_{long}$ was found to depend on 
$\Delta t$ and $m_T$ \cite{garz}.
Here $m_T$ is defined as the average transverse mass given by
$$m_T =\frac{1}{2}\left (\sqrt{m^2+p_{T1}^2}+\sqrt{m^2+p_{T2}^2}\right
)\ ,$$
where $p_{T1}$ and $p_{T2}$ are the transverse momenta of 
the two identical particles \cite{review}. 
The second proposed approach \cite{bialas} rests mainly on the 
Bjorken-Gottfried relation \cite{gottfried}
which associates $R_{long}$ and $R_T$, the longitudinal and transverse dimensions, with 
$m_T$.\\ 
 
Here our study is confined to the first 
approach with the  
aim to gain further insight to 
the BEC and FDC time scale $\Delta t$ associated with \RI.  
In addition to the $Z^0$ hadronic decays
we also examine the BEC results of the identical
hadron pairs produced in central heavy ion collisions. Finally we
draw the attention to the interest in analyzing the BEC
of the $Z^0Z^0$ di-gauge bosons expected to be produced copiously in the
recently constructed 14 TeV $pp$ collider, the LHC at CERN,
in its future high luminosity setup.

\section{Physics background} 
\begin{figure}[h] 
\centering{\psfig{file=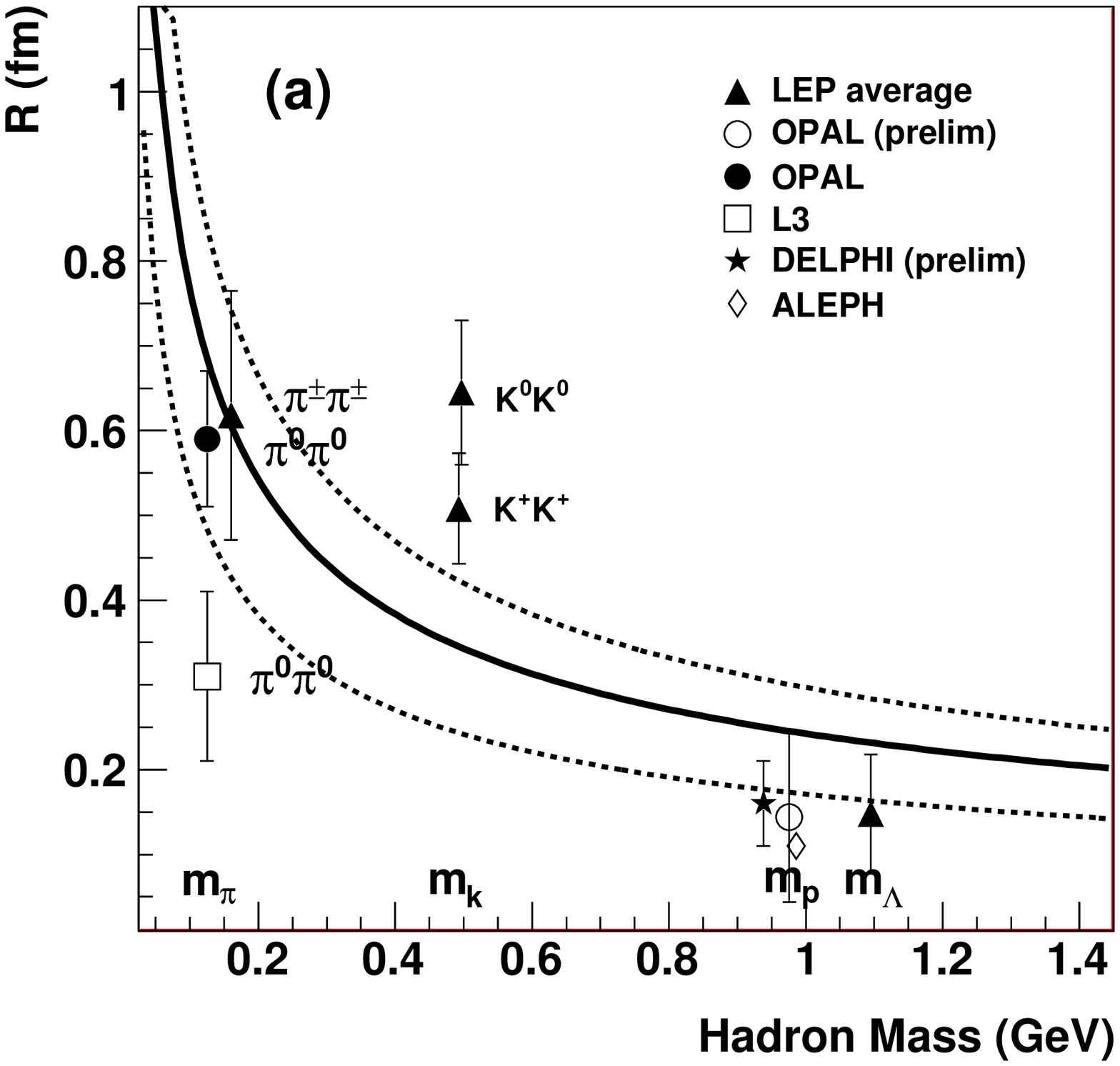,height=8.2cm} 
{\psfig{file=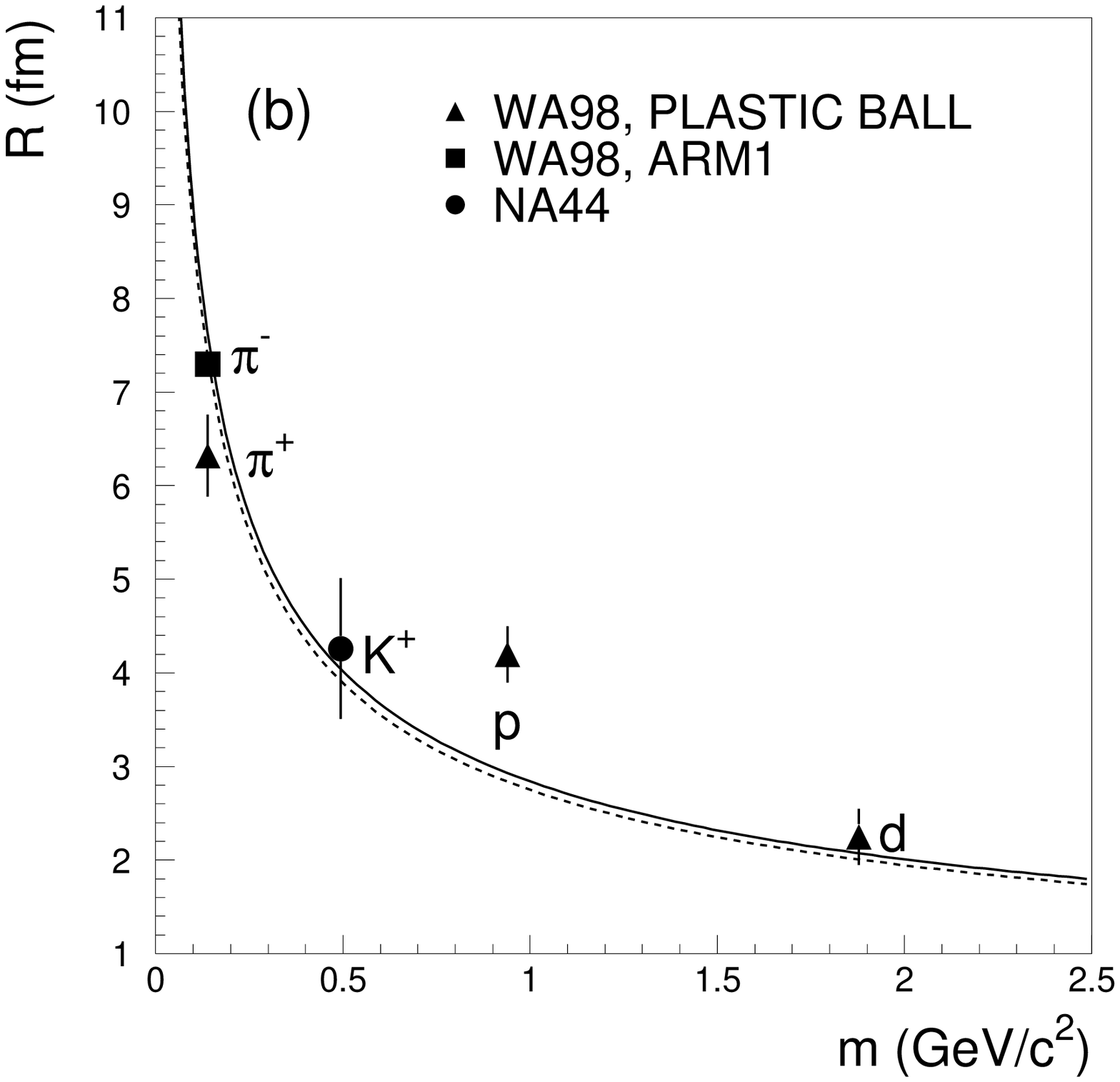,height=7.5cm}}} 
\caption{\small \RI as a function of 
the hadron mass obtained from BEC and FDC analyses. 
(a) Values obtained from the $Z^0$ hadronic decays by the LEP 
experiments \cite{opalpi,l3pi,delphi,alephkk,alephpipi}. The solid 
line represent Eq. 
(\ref{formula}) with $\Delta t$=$10^{-24}$ seconds while 
the dotted lines are for 
$\Delta t$=$(1\pm 0.5)\times10^{-24}$ seconds. 
(b) BEC analyses of hadron pairs 
emerging from central $Pb+Pb$ collisions at 158/A GeV taken
from \cite{wa98}.  
The continuous line is the result of a fit of
Eq. (\ref{formula}) to the data of the Plastic Ball 
detector whereas the dotted line is the result of the fit 
to all the data points shown in the figure.}
\label{wa98mass}
\end{figure}

A compilation of the measured 
one dimension \RI values, 
obtained from BEC and FDC studies of identical hadron pairs 
present in the $Z^0$ decays at LEP, are shown in
Fig. \ref{wa98mass}a
as a function of the outgoing hadron mass $m$. The error bars attached to
the \RI values include the statistical and systematic uncertainties. 
To note is the 
significant spread of the $R_{1D}(m_{\pi^0})$ values between 
two of the experiments. This spread occurs often in the BEC and FDC 
measurements which can be traced back to   
the different adopted procedures and choices of the reference 
sample. Notwithstanding 
this deficiency, \RI is seen to decrease with the increase of
the particle mass. 
That this behavior of $R_{1D}(m)$ is not only limited to the 
$Z^0$ hadronic decays is demonstrated in Fig. \ref{wa98mass}b. 
This figure shows the 
\RI results obtained from
BEC 
and FDC analyses of the outgoing hadrons, from pion to deuteron pairs,  
produced in the central $Pb+Pb$ collisions at 158/A 
GeV \cite{wa98}.\\ 

In Ref. \cite{acl} it has been shown that from 
the Heisenberg uncertainty relations one can derive 
a connection between \RI and the non-zero particle mass 
$m$, namely   
\begin{equation} 
R_{1D}(m)\ =\ \frac{c\sqrt{\hbar \Delta t}}{\sqrt{m}}\ , 
\label{formula} 
\end{equation} 
where $\Delta t$ is a time scale parameter. 
This time scale has been taken in \cite{acl} to be equal to 
$10^{-24}$ seconds representing 
the strong interactions, and thus independent 
of the hadron mass and its identity. As a result, 
the \RI behavior 
on the hadron mass was fairly well reproduced 
by Eq.~(\ref{formula}) which is represented by the continuous line in  
Fig. \ref{wa98mass}a. Furthermore, this line   
essentially coincides with the one 
deduced from a general QCD potential \cite{acl}. 
Here we   
point out that the data given in the figure are the decay product 
of the $Z^0$ gauge boson, the lifetime of which is of the 
order of 
$10^{-24}$ seconds. Thus the 
success of the choice of $\Delta t$=$10^{-24}$ seconds, 
may in fact be, as we further argue,  due to the decay particles' 
emission time which is prescribed by the $Z^0$ boson lifetime.

\section{The time scale $\Delta t$}  
\begin{figure}[ht]
\centering{\psfig{file=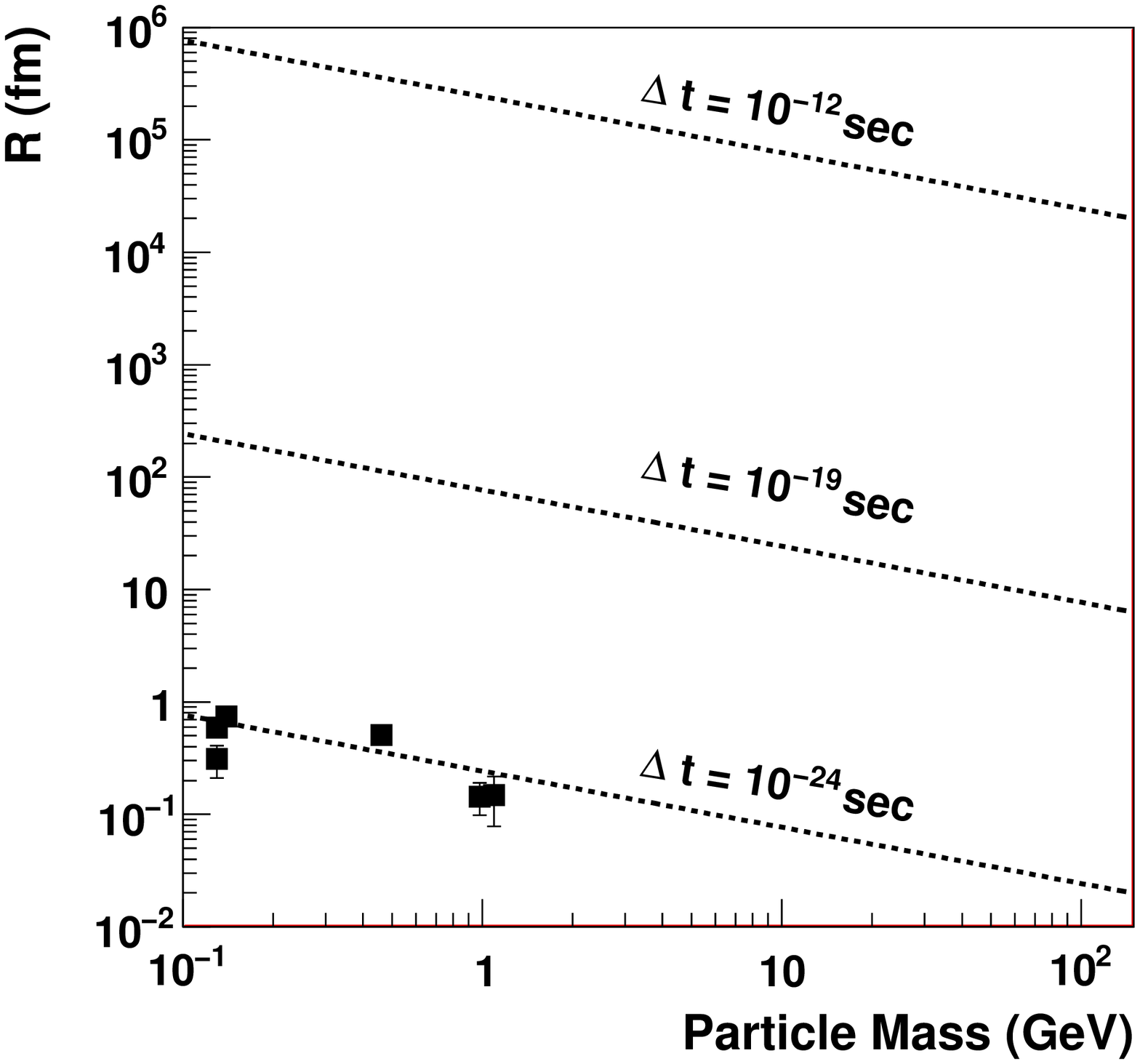,height=8.5cm,}}
\caption{\small The expected \RI  
dependence on the particle mass for 
three $\Delta t$ values of $10^{-24}$, $10^{-19}$ 
and $10^{-12}$ seconds, representing respectively strong, 
electro-magnetic and weak interactions of the outgoing particles.  
The data in the figure are the LEP
measured \RI values from $Z^0$ decays.} 
\label{deltat} 
\end{figure}

It is clear  
that the $R_{1D}(m_{\pi})$ values obtained from 
BEC analyses of pion pairs emerging from collisions like the 
electron-nucleon \cite{zeus} and neutrino-nucleon \cite{nomad} 
are very similar to those obtained in $e^+e^-$ and $pp$ collisions 
and thus exclude the possibility that $\Delta t$ 
is related to the interaction strength of the incoming particles. 
As for the association of $\Delta t$ with the interaction strength of 
the outgoing identical correlated particles, it is instructive 
to examine Fig. \ref{deltat} where, using Eq. (\ref{formula}), 
the expectation of $R_{1D}(m)$ 
are plotted against $m$ for three $\Delta t$ values of $10^{-24}$, 
$10^{-19}$ 
and $10^{-12}$ seconds standing for strong, 
electro-magnetic and weak interactions. 
As can be seen, if indeed 
$\Delta t$ is representing the interaction strength of the outgoing 
particles then for weak interacting  
particles the \RI measured by BEC or FDC 
should reach unreasonable high values as compared to those measured 
in $Z^0$ hadronic decays.\\

An additional evidence against the association of 
$\Delta t$ with the interaction strength of the  
outgoing particles is coming  
from the BEC measurement of the  
non-zero transverse momenta photon pairs, directly produced in 
the central 
$Pb + Pb$ interactions at 158/A GeV \cite{wa98photon}. 
In this case Eq. (\ref{formula}) cannot be applied 
since $m_{\gamma}=0$. However, 
from the Heisenberg uncertainty relations  
one can also derive 
\cite{garz} a relation between the longitudinal 
dimension $R_{long}$, defined in the LCMS, and the average transverse mass  
$m_T$, namely 
\begin{equation} 
R_{long}(m_T)\ =\ \frac{c\sqrt{\hbar \Delta t}}{\sqrt{m_T}}\ , 
\label{formula2} 
\end{equation} 
which is applicable to photons with a non-zero  
transverse momentum. 
The BEC analysis of the directly produced photon pairs of the WA98 
collaboration \cite{wa98photon} 
was  divided into two transverse momentum $P_T$ regions which 
yielded the following \RI and their associated chaoticity 
$\lambda_{1D}$ parameter values:  
$$R_{1D}^I =5.9\pm 1.2\ fm;\ \ \ \lambda^I_{1D}=
0.0028\pm 0.0007;\ \ \ \ {\rm{for}}\ 100<P_T<200\ MeV/c$$
$$R_{1D}^{II} = 6.1\pm 1.4\ fm;\ \ \ \lambda^{II}_{1D}=
0.0029\pm 0.0017;\ \ \ \  {\rm{for}}\ 200<P_T<300\ MeV/c\ ,$$
where the statistical and systematic errors are added in quadrature. 
These $R_{1D}(\gamma)$ values are consistent with the $R_{1D}(m_{\pi})$ 
values obtained in the same $Pb+Pb$ collision experiment at 158/A GeV.
Since in general the $R_{1D}(m)$ values are similar within 10 to 20 $\%$ 
to the corresponding $R_{long}(m_T=m)$, it is also instructive to note that the value 
$R_{1D}(\gamma)\simeq 5.6\ fm$ 
measured at an average $m_T$ of 200 MeV is consistent with the value of 
$R_{long}(m_T=200\ {\rm{MeV}})\simeq 5.9\ fm$ both measured in the $Pb+Pb$
reactions \cite{wa98}.
From this one 
can safely infer that the
$\Delta t$ associated with the directly produced photons is 
in any case 
away by a few orders of magnitude from the $\Delta t$ region that represents  
the elecro-magnetic interaction strength (see Fig. \ref{deltat}). 
Following these observations we assign $\Delta t$ to be the particles' 
emission time.\\ 
 
In the $Z^0$ decay, as well as in hadron interactions,
like in $pp$ reactions, the particle's collision and emission times 
are practically of the same order of magnitude. This apparently
is not the case in heavy ion collisions. 
In the $Pb + Pb$ collisions at 158/A GeV, measured by the
WA98 collaboration \cite{wa98},
the \RI values obtained from   
identical hadron pairs 
are seen in Fig. \ref{wa98mass}b
to be described very well,  
apart from the slight departure of  $R_{1D}(m_p)$, by the continuous 
line. 
This line is the result of a fit of Eq. (\ref{formula}) to the data 
yielding $\Delta 
t=(1.28\pm 0.03)\times 10^{-22}$ seconds, much longer than the
particles' collision time of $\sim$$10^{-24}$ seconds.   
Here it is important to note that the success of this fit is taken by the WA98 collaboration as 
an indication for a common emission duration of the various particle pairs 
produced in the $Pb + Pb$ reactions. Furthermore, in the same
experiment the $R_{long}$ dependence on $m_T$ is very 
well
reproduced by Eq. (\ref{formula2}) with a fit result of
$\Delta t$=$(1.61\pm 0.05)\times10^{-22}$ seconds.\\
 
\begin{figure}[ht] 
\centering{\centering{\psfig{file=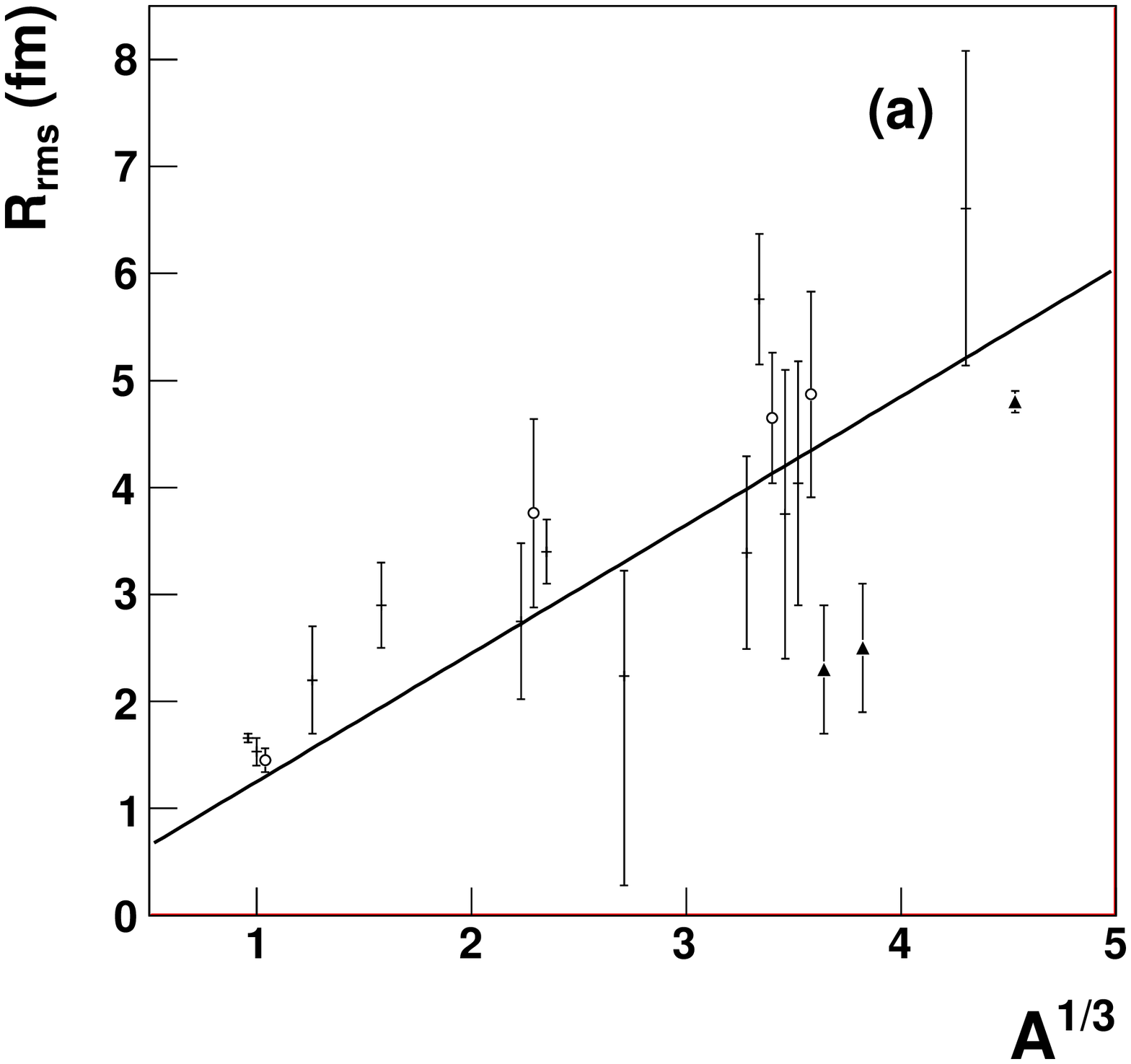,height=7.5cm}}\ \ \ \ 
{\psfig{file=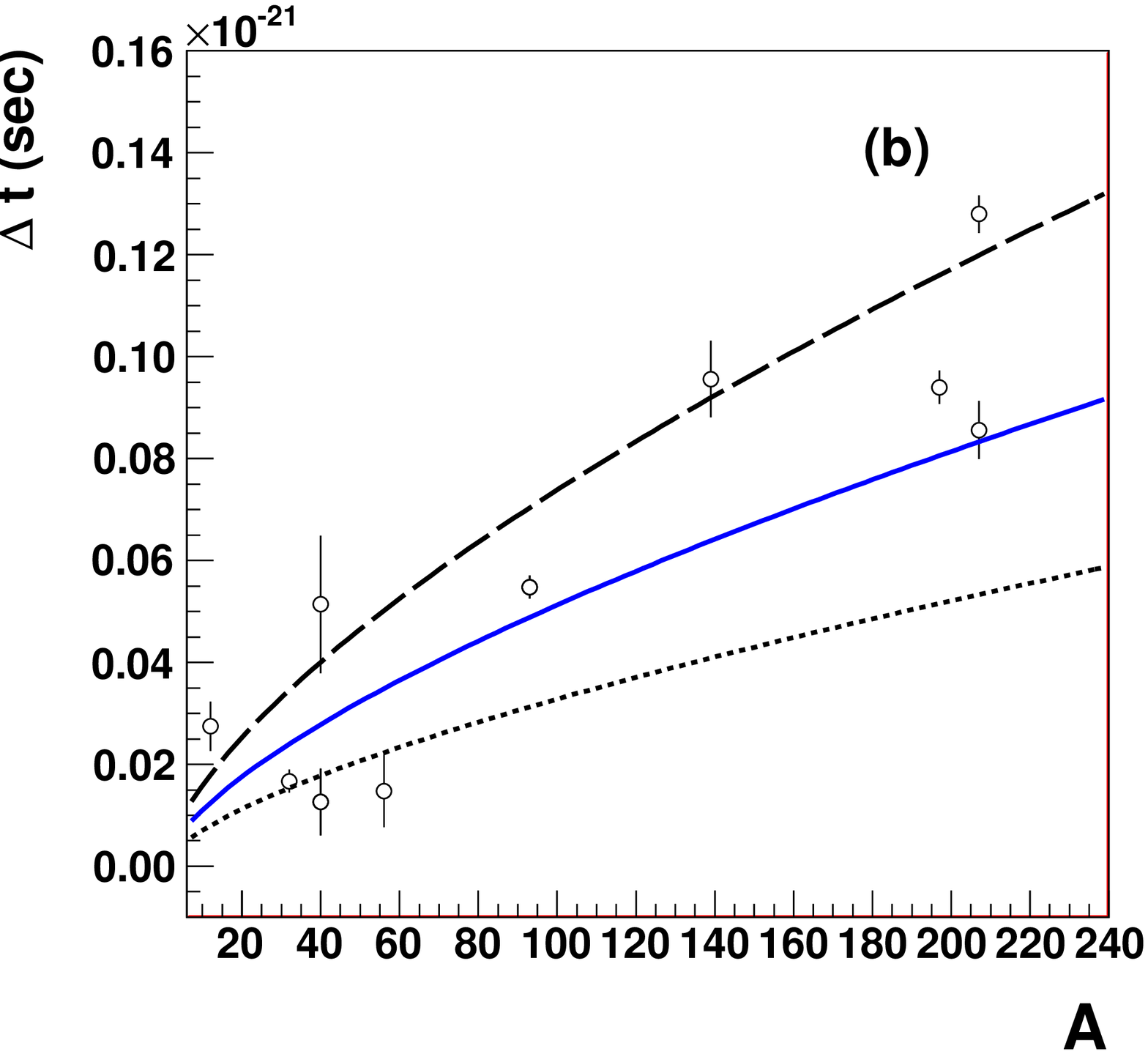,height=7.5cm}}}
\caption{\small (a) The measured \RI from BEC of pion 
pairs produced in heavy 
ions collisions as a function of $A^{1/3}$
reproduced from reference \cite{chacon}. The straight line
represents the relation $R=aA^{1/3}$ with
$a=1.2$ $fm$. 
(b) $\Delta t$ as a function of the atomic number $A$.
The data points are extracted via 
Eq. (\ref{formula})  from the measured \RI values
reported in Refs. \cite{mmmm,LBL+NA44,chacon,wa98inv}.
The dotted, continuous and dashed lines represent 
Eq. (\ref{couple}) respectively for  
$a=0.8,\ 1.0$ and $1.2\ fm$. 
} 
\label{rvsm} 
\end{figure}

Inasmuch that Eq. (\ref{formula}) can also describe successfully 
the $R_{1D}(m)$ results of other heavy ion reactions, apart 
from the $Pb+Pb$ collisions, 
one is naturally led to combine Eq. (\ref{formula}) evaluated
at $m=m_{\pi}$,  
that is
\begin{equation}
R_{1D}(m_{\pi})\ =\ \frac{c\sqrt{\hbar \Delta t}}{\sqrt{m_{\pi}}}\ , 
\label{formulapi}
\end{equation}
with the known 
expression (see e.g. \cite{review}) that relates 
the measured $R_{1D}$($m_{\pi}$) of
pion pairs with the atomic 
number $A$ (see Fig. \ref{rvsm}a), namely 
\begin{equation}
R_{1D}(m_{\pi})=aA^{1/3}\ , 
\label{rvsa}
\end{equation}
where $a$ is a constant of the order of $1.0$ $fm$. 
From these two last equations one obtains 
\begin{equation} 
\Delta t\ =\ \frac{m_{\pi}a^2}{\hbar c^2}A^{2/3}\ . 
\label{couple} 
\end{equation}
Moreover, BEC analyses of pion pairs emerging from central $Pb+Pb$ 
collisions at 
the energies
20/A, 30/A, 40/A, 80/A and 158/A GeV show that the 
radii  
values measured in the LCMS reference frame 
are only very little, if at all,  
dependent on the collision energy \cite{NA49}.
As a consequence, Eq. (\ref{couple}) is expected 
to be 
independent of the collisions energy so that 
$\Delta t$ is essentially only proportional 
to $A^{2/3}$, the surface area of the nucleus.\\  
 
The relation between $\Delta t$ and the atomic number A is presented 
in Fig. \ref{rvsm}b 
where the dotted, continuous and dashed lines are calculated 
from Eq. (\ref{couple}) respectively for 
$a=0.8,\ 1.0\ {\rm{and}}\ 1.2\ fm$. 
The data used for the figure were taken from the well measured 
\RI values of central heavy ion collision experiments, in the atomic number 
range of 12 to 207, 
reported in Refs. \cite{mmmm,LBL+NA44,chacon,wa98inv}. 
From these $R_{1D}$ data the $\Delta t$ values shown in the figure were 
calculated using  
$\Delta t=m_{\pi}R_{1D}^2(m_{\pi})/(\hbar c^2)$ 
as derived from Eq. (\ref{formula}). 
As can be seen, the $\Delta t$ data 
are accounted for by Eq. (\ref{couple}) in or near the range 
defined by 
$a = 0.8\ {\rm{and}}\ a = 1.2\ fm$ lines. 
To note is that the spread of the $\Delta t$ values 
increases, as expected, with $A$. 
As seen from Fig. \ref{rvsm}b  
the common particle emission time is 
predominantly proportional to the nucleus surface area. 
These particle emission times, in the range of $A$ 
covered by Fig. \ref{rvsm}b,  
are of the order of 10$^{-23}$ 
seconds to be compared with the typical estimates  
inferred from 
hydrodynamical-like and other models applied to heavy ion 
collisions \cite{lisa}.\\ 
 
\section{Summary and conclusions}
The role of the time scale $\Delta t$ parameter, associated with
the BEC and FDC of identical particle pairs, has been examined in the 
framework of the Heisenberg uncertainties relations.
From the studies of the $Z^0$ decays, 
the $\Delta t$ is seen to be associated with the particle pair
emission time of the order of $10^{-24}$ seconds, as
determined from the $Z^0$ lifetime, 
rather than being a measure of the particles
interaction strength property.\\ 

In heavy ion collisions one should differentiate
between the typical collision time of $\sim$$10^{-24}$ seconds for 
strong 
interactions, and the emission time of the produced particles. 
This is well illustrated by the WA98 experiment 
where the collision  time is negligible in comparison 
to the particle's emission time of about $10^{-22}$ seconds.\\ 
 
Merging the known dependence of 
$R_{1D}(m_{\pi})$ on the atomic number $A$ namely, 
$R_{1D}(m_{\pi})=aA^{1/3}$, with the \RI dependence on the particle mass 
as derived from the Heisenberg uncertainties, yields the 
equation   
$$\Delta t\ =\ \frac{m_{\pi}a^2}{\hbar c^2}A^{2/3}\ .$$ 
This expression relates the particle emission time $\Delta t$ 
with the surface area 
of the nucleus which agrees with the experimental 
results as is illustrated in Fig. \ref{rvsm}b.\\ 
 
Even though the BEC analysis of two directly produced photons 
in $Pb+Pb$ collisions supports the notion that $\Delta t$ is 
the particle emission time, 
a decisive answer to this issue 
should come from BEC and/or 
FDC of weak interacting particles. 
Presently no such information exists. 
The $\mu^{\pm}\mu^{\pm}$ pairs are in general 
the decay product of pions and/or kaons 
so that they are not produced 
simultaneously. 
As for the $e^{\pm}e^{\pm}$ system 
produced in particle reactions, it also has similar drawbacks. 
For this reason it may be worthwhile to consider 
a BEC analysis of the two weakly interacting $Z^0Z^0$ 
system even though they are expected at lower order to be dominated  
by a coherent production in $pp$ collisions. 
High order corrections may well 
introduced a small non-coherent contributions which will be 
sufficient to allow a BEC analysis as was the case in 
the di-photon BEC analysis \cite{wa98photon} which succeeded 
with a chaoticity $\lambda$ value as small as $\approx 0.003$. 
In case that $\Delta t$ is determined by the 
interaction strength of the $Z^0$ gauge boson then 
its \RI value should be very high as seen from 
Fig. \ref{deltat}. Finally we note that 
the opportunity to carry out a BEC analysis of the $Z^0Z^0$ pair 
may be realized at the 14 TeV CERN Large Hadron $pp$ Collider in its 
upgraded luminosity configuration where sufficiently 
high event statistics can be accumulated.  
 
\subsection*{Acknowledgments} 
We would like to thank E. De Wolf, W. Kittel, E. Levin, M. Lisa, 
W.J. Metzger and E. Sarkisyan-Grinbaum for many helpful 
comments and remarks. Part of this work has been done by G.A. in 
the DESY/Zeuthen laboratory to which he is thankful. 
 
\listoffigures 
\end{document}